\documentclass[useAMS,usenatbib]{mn2e}
\usepackage{graphicx,graphics}

\title[The complex PA swings of radio pulsars]{The complex
  polarization angles of radio pulsars: orthogonal jumps and
  interstellar scattering} \author[A.~Karastergiou]
{A. Karastergiou\\
  Astrophysics, University of Oxford, Denys Wilkinson Building, Keble
  Road,
  Oxford OX1 3RH, UK\\
}\date{\today}
\begin{document}

\date{submitted 28 August 2008}

\pagerange{\pageref{firstpage}--\pageref{lastpage}} \pubyear{2008}

\maketitle

\label{firstpage}

\begin{abstract}
  Despite some success in explaining the observed polarisation angle
  swing of radio pulsars within the geometric rotating vector model,
  many deviations from the expected S-like swing are observed. In this
  paper we provide a simple and credible explanation of these
  variations based on a combination of the rotating vector model,
  intrinsic orthogonally polarized propagation modes within the pulsar
  magnetosphere and the effects of interstellar scattering. We use
  simulations to explore the range of phenomena that may arise from
  this combination, and briefly discuss the possibilities of
  determining the parameters of scattering in an effort to understand
  the intrinsic pulsar polarization.
\end{abstract}


\section{Introduction}
Polarization properties of radio pulsars which are observed in many
sources should form the basis of any attempted interpretation. First
of all, the degree of polarization in pulsars is generally high: often
over 50\% in its linear component (hereafter L) and 10-15\% in its
circular component (hereafter V) (e.g. Gould \& Lyne
1998\nocite{gl98}). Linear polarization decreases with observing
frequency, and is generally low above a few GHz. An exception to this
rule is found in young pulsars with a large spin down energy
derivative $\rm{\dot E}$, which remain highly polarized up to
frequencies above 5~GHz (von Hoensbroech et al. 1998, Weltevrede \&
Johnston 2008 in prep.)\nocite{hkk98,wj08}.

\begin{figure} 
\includegraphics[width=0.49\textwidth]{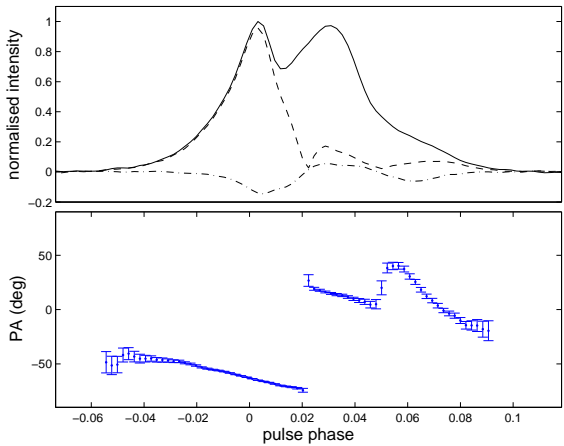}
\caption{\label{psr0355+54} The polarization profile of PSR B0355+54
  from Gould \& Lyne (1998) at 1.4~GHz.  The solid line
  shows the normalised total power, the dashed line the linear
  polarization and the dot-dashed line the circular polarization. The
  issues discussed in this paper are exemplified by the PA in the
  lower panel, which features an orthogonal and non-orthogonal jump
  distorting an apparently smooth swing. }
\end{figure}

In some pulsars, the polarization position angle (hereafter PA),
swings in a smooth way across the pulse profile. The smoothness of the
PA swing lends support to the single rotating vector model of
Radhakrishnan \& Cooke (1969)\nocite{rc69a}, whereby the angle of
polarization is tied to the magnetic field lines, and changes
gradually as the line-of-sight intersects different field lines at
different angles, as:
\begin{equation}
\tan{(PA-PA_0)}=\frac{\sin({\phi-\phi_0})\sin{\alpha}}{\sin{\zeta}\cos{\alpha}-\cos{\zeta}\sin{\alpha}\cos{\phi}}\label{rvm}
\end{equation}
PA$_0$ and $\phi_0$ are constant offsets in PA and phase, $\phi$ the
pulse phase, $\alpha$ the inclination angle between magnetic and
rotation axis, and $\zeta$ the sum of $\alpha$ and the impact
parameter $\beta$. This equation is derived with the convention that
the position angle increases clockwise on the sky. This purely
geometric model has been used to derive the geometry of a number of
pulsars where fitting of PA data is possible (e.g. Everett \& Weisberg
2001)\nocite{ew01}. However, kinks, wiggles and jumps of various
magnitudes often occur in observed PA swings (recent examples can be
found in Johnston et al. 2008)\nocite{jkmg08}. Figure \ref{psr0355+54}
shows the well-known bright pulsar PSR B0355+54 at 1.4~GHz; the data
are from Gould \& Lyne (1998)\nocite{gl98}. The PA swing shows an
abrupt orthogonal jump at pulse phase 0.02, and a bump-like feature at
pulse phase 0.05, superposed on a gradual, S-shaped swing.

In addition, and contradictory to the predictions of the geometric
model, PA swings have been observed not to be entirely frequency
independent (e.g. Karastergiou \& Johnston 2006)\nocite{kj06}. One
known reason for this is that orthogonal polarization modes have been
observed to have different spectral indices (Karastergiou et al. 2005,
Smits et al. 2006), \nocite{kjm05,sse+06} and therefore orthogonal
jumps in the PA swing appear at different locations within the pulse
at different frequencies. A second known frequency dependence of PA
swings is caused by interstellar scattering. Li \& Han
(2003)\nocite{lh03} demonstrated that it is possible to explain low
frequency PA swings by convolving the Stokes profiles of high
frequency data with appropriate scattering functions. They show how
scattering flattens the S-shaped PA curve of the rotating vector
model, rendering the derived geometrical parameters from fitting
erroneous. This was recently also shown for PSR J0908-4913 by Kramer
\& Johnston (2008)\nocite{kj08}, who demonstrated that the PA swing is
indeed independent of frequency if scattering is taken into account.

In this paper, we investigate how observed PA swings like the example
of Figure \ref{psr0355+54} can result from simple intrinsic PA swings,
orthogonal polarization mode jumps and small amounts of scattering. We
show with simulations how orthogonal PA jumps can significantly
distort the PA swing. We demonstrate the modest magnitude of
scattering necessary to explain the observed shapes of PA swings with
orthogonal jumps and study other consequences of this unexplored,
frequency dependent phenomenon.

\vspace*{-0.5cm}
\section{Description of the numerical simulations}

Extensive simulations were carried out to explore scattering effects
on polarization pulse profiles. In contrast to Li \& Han, we use
simulated pulsar data adhering to the rotating vector model for the
PA, rather than high frequency profiles. By examining the scattered
simulated profiles and comparing them to real pulsar profiles, it is
possible to test the effects of interstellar scattering. The software
developed for this purpose follows these steps:
\begin{enumerate}
\item A realistic average pulse profile is generated out of a small
  number of Gaussian components, where the PA follows the rotating
  vector model interrupted only by possible 90$\degr$ jumps.
\item All 4 ``intrinsic'' Stokes parameters are then convolved with
  the desired response function $g(t)$ chosen to simulate the effects
  of scattering and assuming the scattering screen introduces only
  delays and not rotations in the polarization. We have tested three
  such functions,
\begin{eqnarray}
&&g_{ts}(t)=e^{-\Delta t/\tau_s}\label{expo} \\
&&g_{ths}(t)=\left(\frac{\pi\tau_s}{4t^3}\right)^{1/2}\ e^{-\pi^2\tau_s/16t}\label{thick}\\
&&g_{um}(t)=\left(\frac{\pi^5\tau_s^3}{8t^5}\right)^{1/2}\ e^{-\pi^2\tau_s/4t}\label{um}
\end{eqnarray}
which correspond to the simple case of a thin scattering screen at
approximately half way between the source and observer ($g_{ts}$), and
the equations of Williamson (1972)\nocite{w72} for a thick screen near
the source ($g_{ths}$) and a uniform medium ($g_{um}$)
respectively. For a discussion on the merits and shortcomings of
various scattering response functions, see the discussion in Bhat et
al. (2003)\nocite{bcc03}. In all three equations, the scattering
constant $\tau_s$ is the parameter that determines the magnitude of
the effect.
\item The previous step is repeated to simulate an observing backend,
  with a particular number of frequency channels and channel
  bandwidth. For each channel, $\tau_s$ is computed at the center
  frequency, by the empirical equation given in Bhat et
  al. (2004)\nocite{bcc+04}
\begin{eqnarray}
  log(\tau_s)=-6.46+0.154*log(DM)+\nonumber \\
+1.07*log(DM)^2-3.86*log(\nu),\label{tau}
\end{eqnarray}
where $\tau_s$ is in ms, $DM$ is the dispersion measure in cm$^3$pc
and $\nu$ the observing frequency in GHz. Note that observationally
there is quite a large scatter in the measured $\tau_s$ for a given
$DM$. The $DM$ values are chosen to result in small values of
$\tau_s$, generally comparable to the resolution of the simulated
profiles. The exact frequency dependence of $\tau_s$ is far from
totally certain (e.g. L\"ohmer et al. 2004)\nocite{lmg+04}. However,
the analysis presented in the following does not expand to broad
frequency ranges but is confined to typical narrow observing
bandwidths: small changes in the frequency exponent, will incur
relatively small changes in $\tau_s$. A small amount of un-correlated
Gaussian noise is added to each of the Stokes profiles to simulate
real observational data.
\item The average profile is generated by summing up the simulated
  frequency channels.
\end{enumerate}

The degree to which a profile is affected by the convolution process
depends on the relationship between $\tau_s$ and $t_{\rm samp}$, the latter
being the temporal resolution of the data. An interesting handle on
the process is therefore the ratio between the two, $R=\tau_s/t_{\rm samp}$.
\vspace*{-0.5cm}
\section{Simulated and scattered pulse profiles}
\subsection{Profiles with a single orthogonal PA
  jump}\label{sec:single}
\begin{figure*} 
\includegraphics[width=0.49\textwidth]{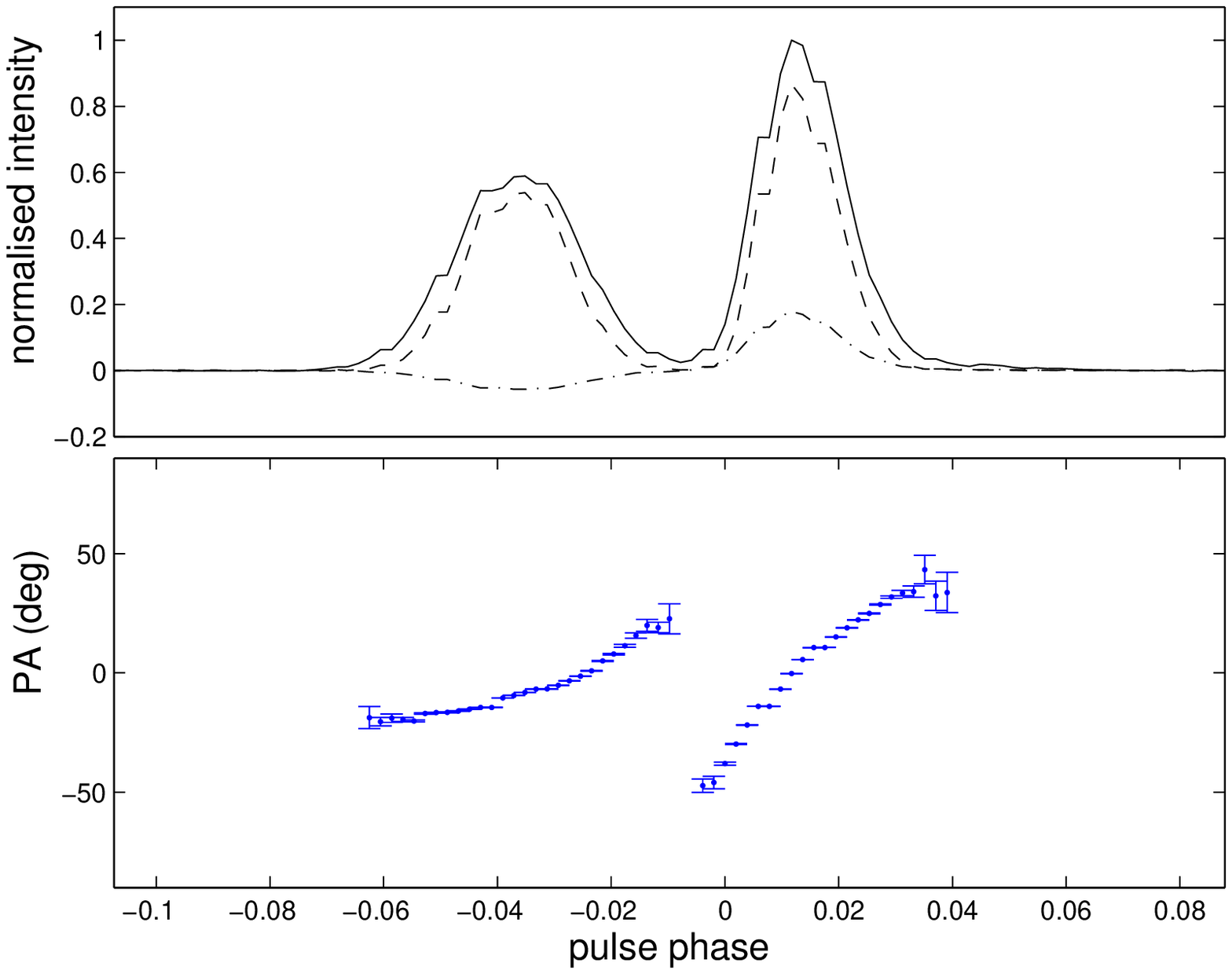}
\includegraphics[width=0.49\textwidth]{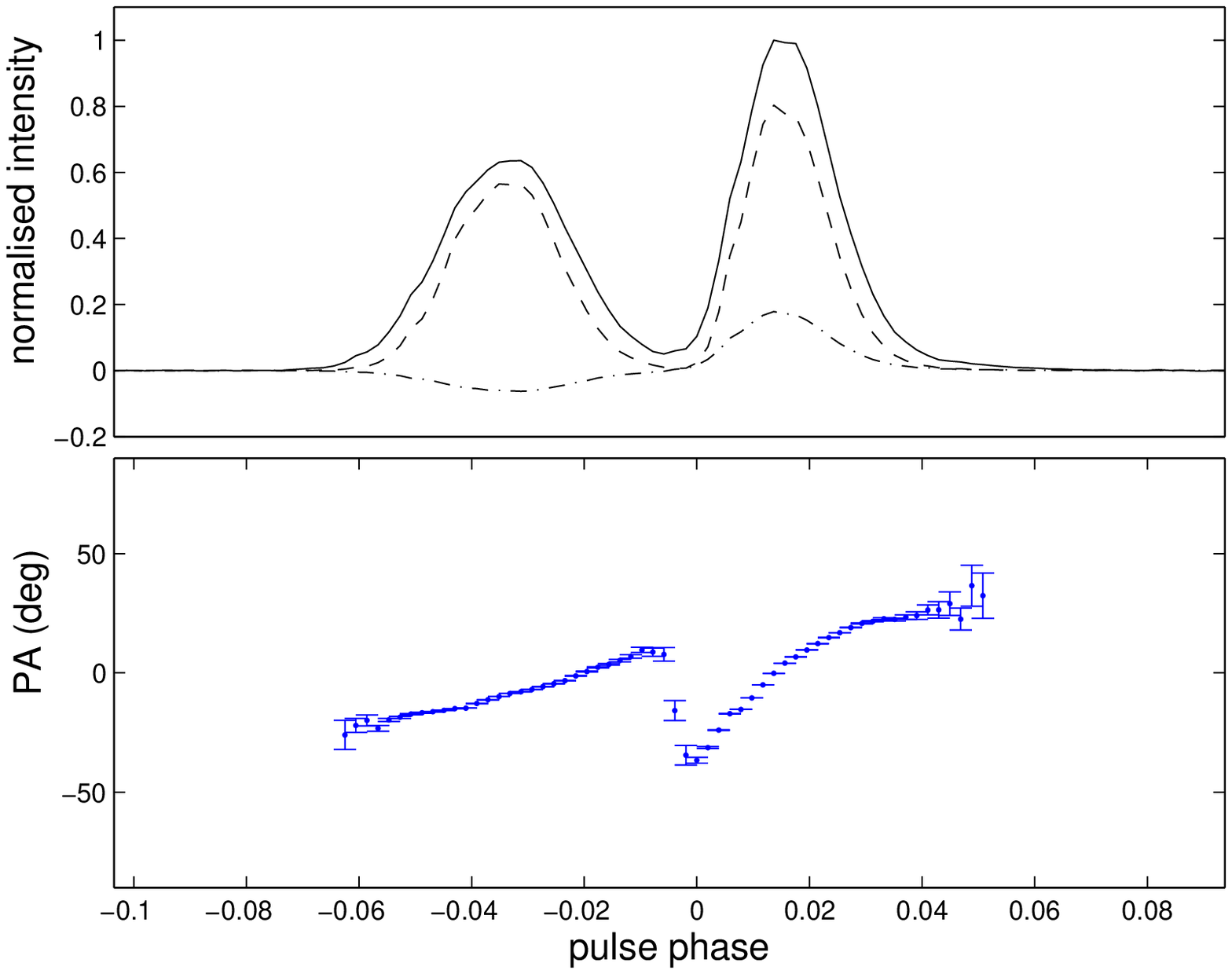}\\
\includegraphics[width=0.49\textwidth]{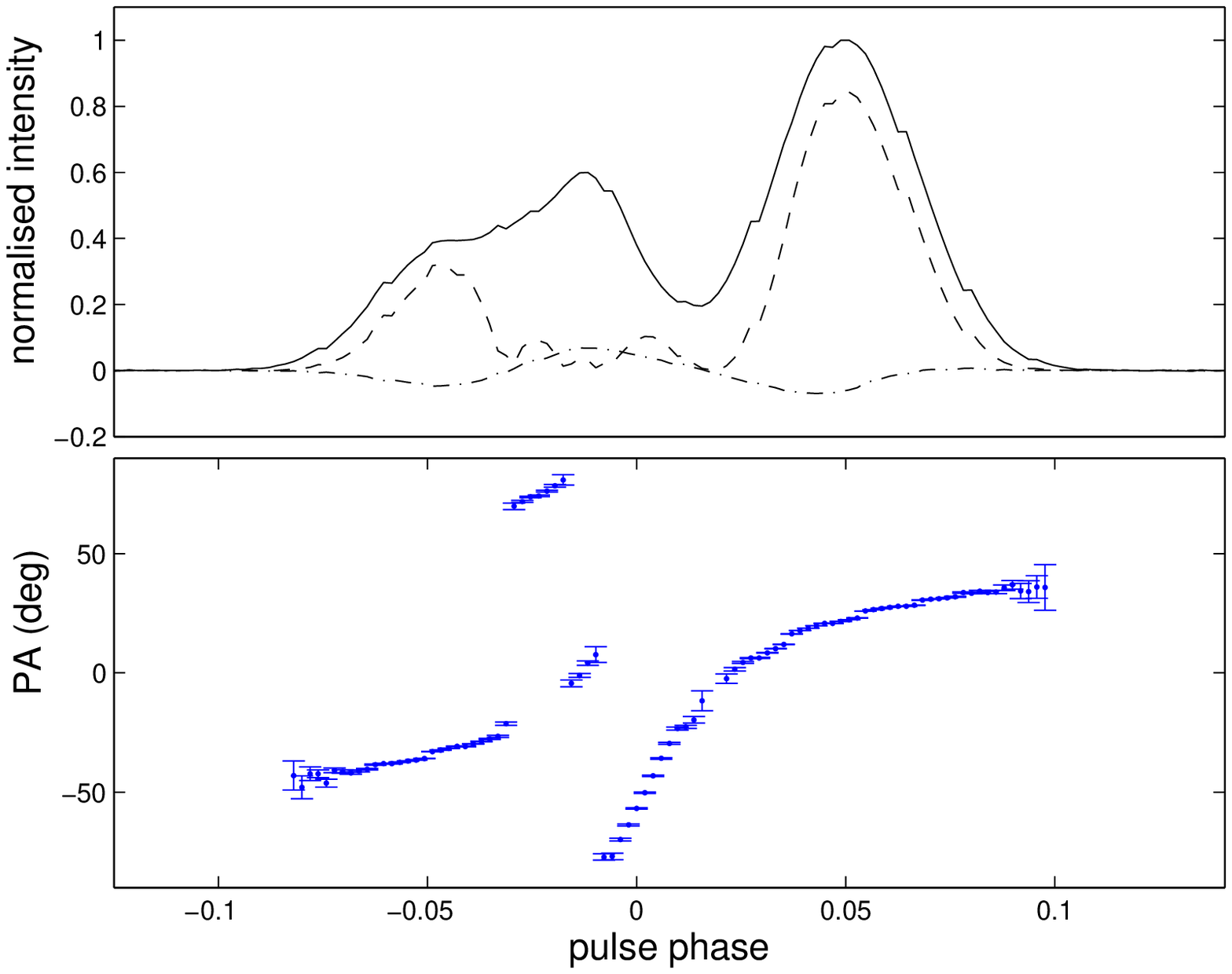}
\includegraphics[width=0.49\textwidth]{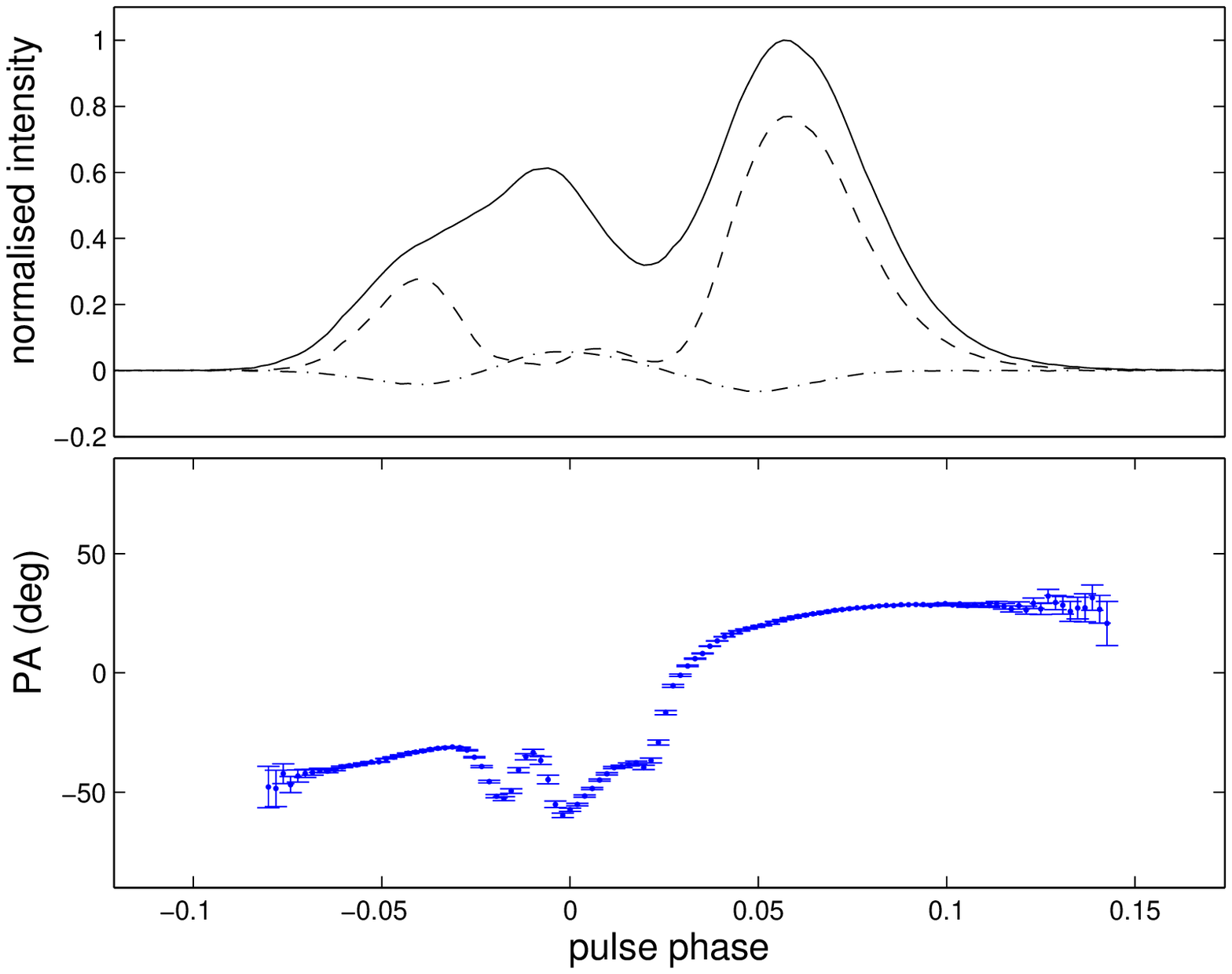}
\caption{\label{figure1}The polarization profile of two simulated
  pulsars (top and bottom row), as in Figure \ref{psr0355+54}. The
  unscattered simulated intrinsic data are on the left, and the
  scattered polarization profiles on the right. The top two plots show
  a 500~ms period pulsar with one orthogonal mode jump (see discussion
  in section \ref{sec:single}), as opposed to a 200~ms pulsar with
  three orthogonal jumps in the bottom plots (see discussion in
  section \ref{sec:multi}).}
\end{figure*}
The top two plots of Figure \ref{figure1} show the input and result of
a typical simulation. The period of the simulated pulsar is 500~ms,
and the profile consists of 512 bins, resulting in a temporal
resolution of $t_{\rm samp}\approx$ 0.976~ms. The ``intrinsic''
profile is shown on the left. The figure on the right demonstrates the
``observed'' profile, with a filterbank of 16 16--MHz wide channels,
centred at 1.4~GHz, assuming equations \ref{expo} and
\ref{tau}. $\tau_s$ ranges between $\sim$2.4 and $\sim$1.25~ms in the
band, based on an assumed value of 340~cm$^3$pc for the $DM$.  The
total power profile (solid line) shows no obvious signs of
scattering. The most obvious consequence of scattering can be seen in
the PA, where the precisely orthogonal jump has flattened out into a
more gradual change of not more that $50\degr$ similar to the real
data in Figure \ref{psr0355+54}. This feature in the PA swing is very
common in real data, and has been a thorn in the side of the model of
orthogonal polarization modes. By applying equation \ref{thick}, a
much more modest $\tau_s$ of order under 1~ms is sufficient to distort
the PA curve in a similar fashion, whereas equation \ref{um} yields
similar results to Figure \ref{figure1} with similar $\tau_s$
values. The degree of linear (dashed line) and circular (dot-dashed
line) polarization remain largely unaffected by this small amount of
scattering.

\vspace*{-0.5cm}
\subsection{Profiles with multiple orthogonal PA
  jumps}\label{sec:multi}
The bottom two plots of Figure \ref{figure1} show another example of a
simulated pulse profile. The pulse period is 200~ms and $t_{\rm
  samp}\approx$0.39~ms. The ``intrinsic'' simulated profile has 3
orthogonal PA jumps, which occur in quick succession between phases
-0.05 and 0. The $\tau_s$ used is the same as the previous example,
ranging from 2.4 to 1.25~ms in the band. Given the shorter pulse
period and the smaller value of $t_{\rm samp}$, the ratio $R$ is
larger ranging from $R\approx 6$ for the lowest frequency channel to
$R\approx 3$ for the highest frequency channel for the same
configuration of filterbank (i.e. 16 16--MHz channels centred on
1.4~GHz). The total power profile does not show evidence of
scattering. However, the PA swing diverges significantly from the
simple geometric model. The kinks and wiggles in the PA swing on the
right are common in real data, and cause severe difficulties both in
fitting the PA data to the rotating vector model, and in theoretical
treatment of the magnetospheric processes.
\begin{figure*} 
\includegraphics[width=1.0\textwidth]{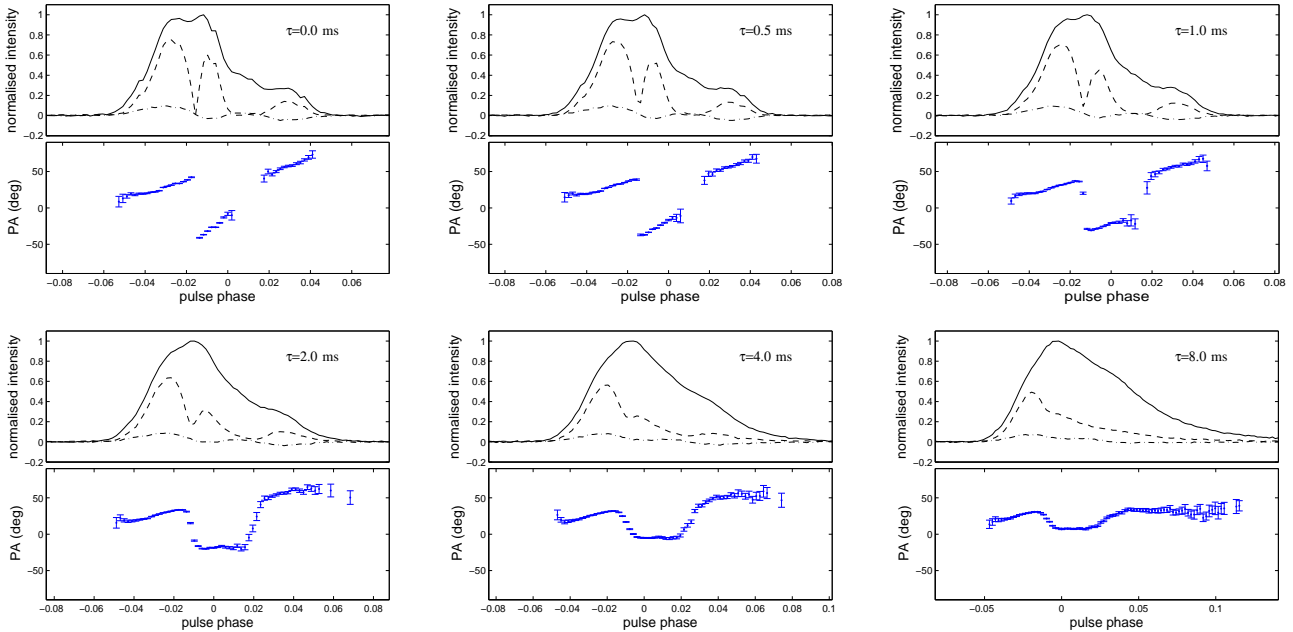}
\caption{\label{strip}The polarization profile of a simulated pulsar
  with a period of 200~ms, scattered by various values of $\tau_s$,
  ranging from 0 to 8~ms as indicated.}
\end{figure*}
Figure \ref{strip} illustrates the way a polarization profile with
orthogonal PA jumps is affected by different values of
$\tau_s$. Again, the profile has a resolution of
$t_{\rm samp}\approx$0.39~ms. Large $\tau_s$ ($\geq$4~ms) values result in
obvious smearing of the total power profile, however the polarization
PA is affected significantly from low $\tau_s$ values. Note how the
steep, middle part of the PA swing, becomes flat at $\tau_s=2$~ms.

\vspace*{-0.7cm}
\subsection{Frequency dependence of the position angle}
\begin{figure} 
\includegraphics[width=1.0\columnwidth]{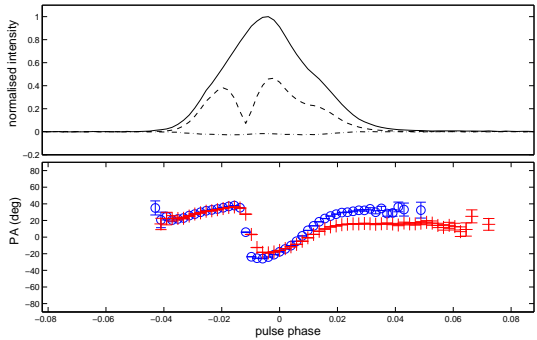}
\caption{\label{figure3}The change in PA across a 256~MHz band
  at 1.4~GHz. Two PA swings are shown: the crosses correspond to the
  lowest and the circles to the highest frequency channel. The
  frequency dependence of $\tau_s$ has quite a pronounced effect even
  at this high observing frequency. }
\end{figure}
\begin{figure} 
\includegraphics[width=1.0\columnwidth]{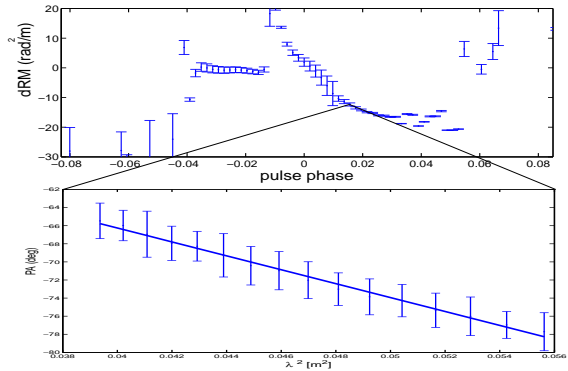}
\caption{\label{figure4}Top: Each bin of the frequency dependent PA
  swing from Figure \ref{figure3} has been fitted for a rotation
  measure in this simulated pulsar, resulting in this phase-resolved
  rotation measure profile. Bottom: The PA of one bin across 16
  frequency channels, plotted versus the wavelength squared. Faraday
  Rotation would result in a linear dependence, which can be fitted
  extremely well here with an $RM$ of -13.42 rad/m$^2$.}
\end{figure}
The fact that small amounts of scattering significantly distort the
pulsar PA swing, has the direct consequence that observed PA curves
are not independent of frequency. This complicates the interpretation
of PA swings, especially in comparing data from multiple, widely
spaced frequencies. Another consequence of scattering manifests itself
within the relatively narrow band of an observation at a given
frequency. Figure \ref{figure3} shows a simulated pulse profile with
an orthogonal PA jump and the PA swings of two frequency channels,
256~MHz apart and centred at 1.4~GHz. The circles correspond to the
highest and the crosses to the lowest frequency. The PA swings are
obviously different, in particular after the pulse phase of the
orthogonal jump. This difference will result in slightly different
polarization profiles for observations with different bandwidth: for
two observations centred on the same frequency, the one with the
largest bandwidth will be more affected.

Pulsars are used to determine Faraday Rotation Measures ($RM$) which
provide information on the Galactic magnetic field in the direction of
the line of sight. $RM$ measurements are based on the assumption that
the only frequency dependence of PA arises from Faraday Rotation
(taking into account potential orthogonal jumps), which, in the
presence of scattering, is not true. A figure showing significant
variation in phase resolved $RM$ measurements across the pulse,
constitutes the basis of an argument on non-orthogonal modes in pulsar
B2016+28 in Ramachandran et al. (2004)\nocite{rbr+04}. The observed
Faraday Rotation is an effect of propagation through the interstellar
medium and not the pulsar magnetosphere, therefore pulse-phase
resolved $RM$ variations are not expected. In that work, an $RM$ is
determined for each pulse phase bin, by fitting the simple equation:
\begin{equation}
\Delta PA=RM\frac{c^2}{\nu^2}\label{RM}
\end{equation}
where $c$ the speed of light and $\nu$ the frequency. 

We considered interstellar scattering as an alternative explanation to
non-orthogonal modes. Simulated scattered data for a band of 16
16--MHz channels at 1.4~GHz were fitted for an $RM$, with no Faraday
Rotation included in the simulation. To match the Ramachandran et al.
analysis, such a fit was carried out for each pulse phase bin (as
shown in the top plot of Figure \ref{figure4}), and although $\tau_s$
depends differently on frequency than the PA due to Faraday Rotation,
the scattered PAs can be very well fitted by equation \ref{RM}. The
results are shown in Figure \ref{figure4}: in the top panel, the phase
resolved fitted $RM$ is shown with errorbars representing the goodness
of the fit. It is denoted as $dRM$ on the axis, as it would appear
additional to the real interstellar $RM$ of a given source. The bottom
panel shows the fit for one phase bin with an apparent negative $RM$
purely due to scattering.  The results from these scattering
simulations resemble the PSR B2016+28 data, and the PA swing of that
pulsar shows evidence of an orthogonal jump (non-orthogonal most
likely due to scattering). It is therefore apparent that modest
scattering should be considered as an alternative, and in many ways
more simple explanation to non-orthogonal modes.

Scattering has implications on the way the $RM$ can be correctly
measured. Figure \ref{figure4} clearly shows that choosing one phase
bin, even if it is the most polarized, and fitting for an $RM$ can
lead to erroneous results. Similarly, any approach which directly
compares PA values of highly polarized bins between channels of
different frequencies will also be inaccurate, as demonstrated in
Figure \ref{figure3}. The only way to avoid issues pertaining to
scattering, is to perform a sum across the profile of the Stokes
parameters Q and U for each frequency channel (Noutsos et
al. 2008)\nocite{njkk08}. As scattering only transfers power into
(mainly) later pulse phases, it should not affect the sum over the
entire profile. The simulations carried out here verify this, and no
matter how distorted the phase resolved $RM$ profile, a calculation of
$RM$ using the proposed method always yields the $RM$ used for the
simulation within the accuracy of the measurement.  

\vspace*{-0.7cm}
\section{Discussion}
We have shown in the previous section that small amounts of scattering
can play an important role in forming the PA swing. We have
concentrated on the frequency of 1.4~GHz, as this is where the bulk of
observational data lie. In Figures 2 and 3, we have demonstrated the
potential of scattering and orthogonal PA jumps to distort the PA
swing. Figures 4 and 5 indicate the difficulties this creates on $RM$
measurements. It is obvious from these figures that the phenomenon
described here depends critically on the presence and location of
orthogonal PA jumps, the degree of scattering and the steepness of the
intrinsic PA swing.

As it has been shown in the past and most recently by Kramer \&
Johnston (2008), in simple PA swings without orthogonal jumps,
scattering makes the PA swing flatter. Similarly, if an orthogonal
jump is located where the PA is intrinsically flat, scattering cannot
create the dramatic kinks and wiggles shown here. The greatest effect
is then achieved when orthogonal PA jumps occur near the steepest
gradient of the PA swing, as given by equation \ref{rvm}. This is an
interesting conclusion, as it is a well established observational fact
that the rotating vector model fails predominantly in components
nearer to the magnetic axis where the gradient of the PA is steepest
(Rankin 1990)\nocite{ran90}. A combination of orthogonal PA jumps in
this central part of the profile and some scattering, has the
potential of generating PA swings only vaguely resembling the simple
predicted S-shaped curve. The effect is even further pronounced if
orthogonal PA jumps occur in very rapid succession, compared to the
scattering constant $\tau_s$. If the assumption is made that each
emitting patch in the patchy radio beam is in a given polarization
mode, a quick succession of jumps near the center of the pulse is an
indication that the central components are narrower, as expected if
they are emitted at lower altitudes above the pulsar surface
(Karastergiou \& Johnston 2007).

The combination of a rapidly swinging intrinsic PA, which rapidly
jumps by $90\degr$, with scattering can then generate a huge variety
of PA swings. As the observed profile is determined by the
combination of these effects, a fitting method for the rotation vector
model to determine the emission geometry should ideally simultaneously
fit for scattering. This will be the object of a more extensive study
on this topic, to be presented in the future.

Scattering through the interstellar medium is a stochastic phenomenon,
and here we only consider integrated pulse profiles. However,
scattering should be similarly responsible for distortions to the
polarization of individual pulses. In particular the distributions of
single pulse PA values in particular bins are often quoted to be
broader than expected by the presence of instrumental noise
(e.g. McKinnon 2004, Karastergiou et
al. 2002\nocite{mck04,kkj+02}). The next version of scattering
simulations will examine the degree to which scattering can reproduce
aspects of the observed pulse-to-pulse PA phenomenology.
\vspace*{-0.6cm}
\section{Concluding remarks}

We have demonstrated that orthogonal jumps in the PA swings of
pulsars, together with very modest amounts of scattering, can lead to
distorted PA swings similar to those observed (see Gould \& Lyne 1998,
Everett \& Weisberg 2001 and others for many examples). The natural
question that arises is how the effect of scattering can be taken out
of the observed data, to recover the intrinsic polarization. Bhat et
al. (2003) proposed a scheme, based on the CLEAN algorithm, to attempt
this, with considerable success. Attempting this on the full Stokes
parameter data seems imperative. Polarization data may impose
limitations to the scattering response function and the scattering
constant, permitting only orthogonal jumps of PA and an intrinsic PA
curve strictly following the rotating vector model in equation
\ref{rvm}. This area warrants significant further attention, in an
ongoing effort to understand the physics of pulsar radio emission.
\vspace*{-0.8cm}
\bibliography{journals,modrefs,psrrefs,crossrefs,somerefs}
\bibliographystyle{mn2e}
\label{lastpage}

\end{document}